\documentclass[aps,prd,twocolumn,superscriptaddress,nofootinbib,10pt]{revtex4-2}
\usepackage[paperwidth=21cm,paperheight=29.7cm,top=2.54cm,bottom=2.54cm,left=2cm,right=2cm]{geometry}
\usepackage[colorlinks,linkcolor=blue,anchorcolor=blue,citecolor=blue,urlcolor=blue]{hyperref}
\usepackage{graphicx,subfigure}
\usepackage[figuresright]{rotating}
\usepackage{amsmath,amsfonts,amssymb,bm}
\usepackage{array,enumitem,multirow}
\usepackage{acronym}
\usepackage{makecell}
\newcommand{\be}{\begin{equation}}
\newcommand{\ee}{\end{equation}}
\newcommand{\bea}{\begin{eqnarray}}
\newcommand{\eea}{\end{eqnarray}}
\newcommand{\nn}{\nonumber}

\newcommand{\TQC}{MOE Key Laboratory of TianQin Mission, TianQin Research Center for Gravitational Physics \&  School of Physics and Astronomy, Frontiers Science Center for TianQin, Gravitational Wave Research Center of CNSA, Sun Yat-sen University (Zhuhai Campus), Zhuhai 519082, China.}

\newacro{GR}{general relativity}
\newacro{GW}{gravitational wave}
\newacro{MG}{modified gravity theory}
\newacro{BH}{Black hole}
\newacro{PN}{post-Newtonion}
\newacro{ppE}{parameterized post-Einsteinian}
\newacro{GCB}{galactic ultra-compact binary}
\newacro{SBHB}{stellar-mass black hole binary}
\newacro{MBHB}{massive black hole binary}
\newacro{BHB}{black hole binary}
\newacro{IMBHB}{intermediate-mass black hole binary}
\newacro{EMRI}{extreme mass ratio inspiral}
\newacro{IMRI}{intermediate mass ratio inspiral}
\newacro{SGWB}{stochastic gravitational wave background}
\newacro{MECO}{minimal energy circular orbit}
\newacro{FAR}{false alarm rate}
\newacro{CE}{Cosmic Explorer}
\newacro{ET}{Einstein Telescope}
\newacro{LISA}{Laser Interferometer Space Antenna}
\newacro{BBO}{Big Bang Observer}
\newacro{LVK}{LIGO-Virgo-KAGRA}
\newacro{EdGB}{Einstein-dilaton Gauss-Bonnet}
\newacro{dCS}{dynamic Chern-Simons}
\newacro{SNR}{signal-to-noise ratio}
\newacro{FIM}{Fisher Information Matrix}
\newacro{ISCO}{innermost stable circular orbit}
\newacro{NSBH}{neutron star-black hole binary}
\newacro{MCMC}{Markov Chain Monte Carlo}
\newacro{QNM}{quasi-nomral mode}

\begin{document}

\title{Constraining the dynamical Chern-Simons gravity with future gravitational wave detectors}

\author{Xinyi Che}
\affiliation{\TQC}
\author{Xiangyu Lyu}
\affiliation{\TQC}
\author{Changfu Shi}
\affiliation{\TQC}
\email{Email: shichf6@mail.sysu.edu.cn (Corresponding author)}

\date{\today}

\begin{abstract}
Dynamical Chern-Simons gravity, a parity-violating modification of general relativity, is regarded as a low-energy effective theory arising from string theory. Gravitational waves provide a powerful probe for testing its predictions. However, current gravitational wave observations are unable to place meaningful constraints on this theory through phase measurements, due to limitations from detector noise and the validity requirements of the waveform models.
In this paper, we conduct a comprehensive assessment of the prospects for constraining the dynamical Chern-Simons gravity with future gravitational-wave detectors using stellar mass black holes binary. We quantify how the constraining capacities vary across different detectors and source parameters, and identify the regions of parameter space that satisfy the small-coupling  condition. Furthermore, by incorporating an astrophysically motivated mass distribution model for stellar mass black hole binaries, we estimate the potential of upcoming observatories.
\end{abstract}

\maketitle

\section{Introduction}

The detection of \ac{GW} from compact binary has opened the door to study the nature of gravity and dark compact objects in the strong field and dynamical regime. By using the currently available \ac{GW} data \cite{LIGOScientific:2016aoc,LIGOScientific:2018mvr,LIGOScientific:2020ibl,LIGOScientific:2021usb,LIGOScientific:2021djp}, a variety of tests have been performed \cite{LIGOScientific:2016lio,LIGOScientific:2018dkp,LIGOScientific:2019fpa,LIGOScientific:2020tif,LIGOScientific:2021sio,Perkins:2021mhb, Wang:2021jfc,Niu:2021nic,Wang:2021ctl,Kobakhidze:2016cqh,Yunes:2016jcc}, such as  theory agnostic tests \cite{LIGOScientific:2016lio,Ghosh:2016qgn,Arun:2006yw}, the tests on specific topics \cite{Isi:2019aib, Isi:2017equ, LIGOScientific:2016lio,Krishnendu:2017shb, Mirshekari:2011yq,Yunes:2016jcc,Vijaykumar:2020nzc,Niu:2022yhr}, and the tests targeting different modified theories of gravity \cite{Yunes:2016jcc,Perkins:2021mhb,Wang:2021jfc,Niu:2021nic,Wang:2021ctl,Zhao:2019suc,Okounkova:2021xjv,Kobakhidze:2016cqh,Jenks:2020gbt,Zhu:2022uoq,Wu:2021ndf,Wang:2021gqm,Wang:2020cub,Haegel:2022ymk,Gong:2021jgg,Du:2020rlx}. No evidence against \ac{GR} has been found so far.

Among those modified theories of gravity, \ac{dCS} gravity \cite{Jackiw:2003pm} is a well-motivated extension of \ac{GR} that arises naturally in the low-energy effective action of string theory frameworks \cite{Alexander:2004us,Alexander:2004xd}. It introduces a parity-violating correction to the Einstein–Hilbert action by coupling a scalar field to the Pontryagin density, a topological invariant constructed from the Riemann curvature tensor. This coupling modifies the gravitational field equations in a way that breaks parity symmetry, leading to distinctive signatures in the dynamics of spinning compact objects \cite{Yunes:2009hc} and the propagation and generation of GW \cite{Li:2022grj}.
dCS gravity has been constrained using astrophysical \cite{Yagi:2012ya}, solar-system \cite{Nakamura:2018yaw}, and cosmological data. Currently, the
tightest constraint on this theory comes from the multi-message observation of neutron star systems, combining X-ray and GW data, giving $\sqrt{\bar{\alpha}_{\rm dCS}}<8.5$ km \cite{Silva:2020acr}.

In recent years, testing \ac{dCS} gravity with \acp{GW} has attracted significant interest. The foundation for such tests lies in waveform models incorporating dCS corrections. Yunes and Pretorius \cite{Yunes:2009hc} derived the leading-order modification to the gravitational waveform for slowly rotating black holes. Subsequent studies extended these results to higher post-Newtonian orders and included spin effects, enabling accurate modeling of phase shifts during the inspiral phase of binary black hole coalescences—under the small-coupling approximation and for quasi-circular, non-precessing orbits \cite{Yagi:2011xp,Tahura:2018zuq,Li:2022grj}.

Following the direct GW detection by \ac{LVK} Collaboration, these modified templates have been employed to constrain the dCS coupling parameter. Nair et al. \cite{Nair:2019iur} were the first to apply real GW data to test dCS gravity, performing a Bayesian analysis on events from the GWTC-1 catalog \cite{LIGOScientific:2018mvr}. However, their inferred constraints violated the weak-coupling condition required for the validity of the perturbative waveform model, making the resulting bounds physically uninterpretable and thus unable to provide a meaningful test of the theory.
A series of follow-up studies \cite{Perkins:2021mhb,Wang:2021jfc,Lyu:2022gdr,Wang:2023wgv,Silva:2022srr,Chung:2025wbg} have since undertaken more comprehensive investigations. These works incorporated additional GW events, considered black hole–neutron star mergers, included higher-order waveform modes and harmonic corrections, and even accounted for contributions from the ringdown phase. Despite these enhancements, their conclusions remain consistent with \cite{Nair:2019iur}: the derived constraints still fail to satisfy the weak-coupling validity criterion. Consequently, no existing analysis has yet yielded a robust, self-consistent test of dCS gravity using real gravitational-wave observations.

With the commissioning of more GW detectors and ongoing advances in observational capabilities, significantly tighter constraints on \ac{dCS} gravity are expected in the coming years.
In this paper, we investigate the ability of current and future GW observatories to constrain the dCS coupling parameter using phase measurements from the inspiral signals of binary black hole systems. The detectors under consideration include: the operational ground-based detector aLIGO; third-generation ground-based facilities, i.e. \ac{CE} \cite{Reitze:2019iox,Evans:2021gyd} and \ac{ET} \cite{Punturo:2010zz,ET:2025xjr}; space-based missions, TianQin \cite{Luo:2015,Luo:2025ewp,Li:2024rnk}, LISA \cite{Audley:2017drz,Robson:2018ifk}, and Taiji \cite{Hu:2017mde}; as well as sub-Hertz concepts such as DECIGO \cite{Kawamura:2006up} and \ac{BBO} \cite{Harry:2006fi}.
Using the Fisher information matrix formalism, we assess the expected parameter estimation accuracy for each detector, identify the corresponding requirements on binary source properties, and thereby evaluate the feasibility of performing meaningful tests of dCS gravity with these present and future observatories.

The paper is organized as follows. In Sec. \ref{sec:method}, we brief introduce some basic of dCS gravity and the waveform model, the sensitivity of detectors and data analysis methods. In Sec. \ref{sec:result}, we present our main results. A summary is presented in section \ref{sec:conclusion}. In this study, we refer to the natural units in which $G_N=\bar{h}=c=1\,$ unless otherwise specified.

\section{Method}
\label{sec:method}
In this section, we will introduce the main tools used in this paper. First of all, we will give a brief introduction of \ac{dCS} gravity and the \ac{GW} waveform predicted by this theory for black hole binary. Then, we will also introduce the sensitivities of detectors and the statistical methods employed.

\subsection{Theory and waveform}
Inspired by quantum gravity theories such as string theory and loop quantum gravity, \ac{dCS} extends \ac{GR} by introducing a new ingredient into the gravitational action: a dynamical scalar field that evolves in both space and time. This field is coupled to spacetime curvature through a topological term known as the Chern-Simons term, which takes the form of dual Riemann tensor contracted with the Pontryagin density. As a distinguished consequence, this theory explicitly violates parity symmetry in the gravitational sector.

The action of \ac{dCS} gravity can be written as:
\begin{align}
	\label{eq:action}
	S = S_{\text{EH}} + S_{\text{CS}} + S_{\text{scalar}} + S_{\text{matter}},
\end{align}
where $S_{\text{EH}}$  is the standard Einstein-Hilbert action,
$ S_{\text{scalar}} $ describes the kinetic and potential terms for the scalar field,
$ S_{\text{matter}} $ includes all non-gravitational matter fields, and the novel Chern-Simons correction is given by:
\begin{align}
	S_{\text{CS}} = \frac{1}{32\pi G} \int d^4x\, \sqrt{-g}\, \bar{\alpha}_{\rm dCS} \phi\, {}^\ast\!RR
\end{align}
where $ \phi $ is the dynamical scalar field, $\bar{\alpha}_{\rm dCS}$ is the coupling constant, ${}^\ast\!RR = \frac{1}{2} \epsilon_{\mu\nu\rho\sigma} R^{\rho\sigma}_{\ \ \alpha\beta} R^{\mu\nu\alpha\beta} $ is the Pontryagin density, and $ \epsilon_{\mu\nu\rho\sigma} $ is the Levi-Civita tensor.

\ac{dCS} gravity imprints characteristic corrections on the GW waveform. Currently, waveform modifications from modified theories of gravity are reliably computed only for the inspiral phase of compact binary coalescence. During this stage, the components are widely separated and move at velocities much smaller than the speed of light, allowing the dynamics to be accurately described using the \ac{PN} approximation, particularly for systems with comparable masses. Following \cite{Yunes:2009ke}, the frequency-domain PN waveform for the inspiral can be expressed as:
\begin{align}
	\label{eq:pn}
	h(f)&=h_{GR}(1+\alpha u^a)e^{i\beta u^b}.
\end{align}
Here, $h_{GR}(f)$ represent the \ac{GW} waveform as predicted by GR. \(\alpha u^{a}\) and \(\beta u^{b}\) denote the leading-order corrections to the waveform amplitude and phase, respectively. The exponents \(a = 2~\text{PN}\) and \(b = 2~\text{PN} - 5\) indicate the \ac{PN} order at which these corrections first appear. The parameters \(\alpha\) and \(\beta\) quantify deviations from \ac{GR}; in \ac{GR}, both vanish: \(\alpha = 0\) and \(\beta = 0\).
The characteristic velocity of the binary system is defined as
\[
u = \left( \pi M \eta^{3/5} f \right)^{1/3},
\]
where \(M = m_1 + m_2\) is the total mass, \(\eta = m_1 m_2 / M^2\) is the symmetric mass ratio, and \(m_1\) and \(m_2\) are the masses of the primary and secondary components, respectively.
Since laser interferometric GW detectors are significantly more sensitive to phase than to amplitude, this work focuses exclusively on the phase modification.

Following  Tahura et al. \cite{Tahura:2018zuq}, the correction introduced by \ac{dCS} to the \ac{GW} phase for a binary system can be written:
\begin{align}
	\label{eq:dcs}
	\delta\Psi&=\frac{57713\eta^{-14/5}}{344064}\frac{16\pi \bar{\alpha}_{\rm dCS}^2}{M^2}\Big[\Big(1-\frac{14976\eta}{57713}\Big)\chi_a^2\nn\\
	&+\Big(1-\frac{215876\eta}{57713}\Big)\chi_s^2-2\delta_m\chi_a\chi_s\Big](\pi M\eta^{3/5} f)^{-1/3},
\end{align}
where  $\delta_m\equiv(m_1-m_2)/M\,$, $\chi_s=(\chi_1+\chi_2)/2\,$, $\chi_a=(\chi_1-\chi_2)/2\,$, $\xi_{\rm dCS}\equiv16\pi\bar{\alpha}^2_{\rm dCS}/M^4$, \(\chi_1\) and \(\chi_2\) are the dimensionless spins of the primary and secondary components, respectively. From Eq.\eqref{eq:dcs}, one can conclude:
\begin{align}
	\label{eq:ppedcs}
	\beta_{\rm dCS}&=-\frac{57713\eta^{-14/5}\xi_{\rm dCS}}{344064}\Big[\Big(1-\frac{14976\eta}{57713}\Big)\chi_a^2\nn\\
	&+\Big(1-\frac{215876\eta}{57713}\Big)\chi_s^2-2\delta_m\chi_a\chi_s\Big].
\end{align}
Here, \( b_{\rm dCS} = -1 \) signifies that the leading-order correction to the GW waveform induced by dCS gravity appears at the 2PN order. However, to ensure the perturbative well-posedness of the theory, current dCS waveform models are derived under the small-coupling approximation, in which deviations from GR are treated as small perturbations. This assumption is well justified by the remarkable consistency of GR with a wide range of observations, implying that any modification must be subdominant. Consequently, the dCS coupling parameter must satisfy the theoretical validity bound \cite{Perkins:2021mhb, Lyu:2022gdr}:
\begin{align}
	\bar{\alpha}_{\rm dCS}^2 \lesssim \frac{m_2^4}{32 \pi} \,.
	\label{eq:validity}
\end{align}

Equation~\eqref{eq:pn} is valid only when the dominant \((2,2)\) mode is considered. However, for systems with large mass ratios or significant eccentricities, higher-order waveform modes contribute non-negligibly and must be included. A natural generalization of Eq.~\eqref{eq:pn} that accounts for multiple harmonic modes is therefore:
\begin{align}
	h_{\rm dCS}(f) &= \sum_{\ell m} h_{{\rm dCS},\ell m}(f), \\
	h_{{\rm dCS},\ell m}(f) &= h^{\rm GR}_{\ell m}(f) \, \exp\!\left(i \beta_{{\rm dCS},\ell m} \, u^{b_{{\rm dCS},\ell m}} \right),
\end{align}
where \(\beta_{{\rm dCS},\ell m}\) and \(b_{{\rm dCS},\ell m}\) denote the phase correction amplitude and PN order for the \((\ell,m)\) mode, respectively. As shown in \cite{Mezzasoma:2022pjb}, for non-precessing binaries, these parameters scale with the harmonic indices as
\begin{align}
	\label{eq:waveform}
	\beta_{{\rm dCS},\ell m} &= \left( \frac{2}{m} \right)^{b_{\rm dCS}/3 - 1} \beta_{\rm dCS}, \\
	b_{{\rm dCS},\ell m} &= b_{\rm dCS},
\end{align}
with \(\ell = 2,3,4,\dots\) and \(m = 0, \pm1, \pm2, \dots\).
In this work, we use the \texttt{IMRPhenomXHM} \cite{Garcia-Quiros:2020qpx} to model the GW waveform predicted by GR, including the following higher harmonic modes: \((\ell, m ) = \{(2,2), (2,1), (3,3), (3,2), (4,4)\}\).

\subsection{Statistical methods and detector configuration}
We employ the \ac{FIM} method \cite{Finn:1992wt,Cutler:1994ys} to estimate the expected constraints on the dCS coupling constant \(\bar{\alpha}_{\rm dCS}\). Under the assumptions of high \ac{SNR} and stationary Gaussian noise, the statistical uncertainty in each parameter is approximated by
\begin{equation}
	\Delta\theta^a \equiv \sqrt{\langle \Delta\theta^a \Delta\theta^a \rangle} \approx \sqrt{(\Gamma^{-1})^{aa}} \,,
	\label{estimation}
\end{equation}
where \(\langle \cdots \rangle\) denotes an ensemble average.
\(\theta\) represent the parameters that are considered in the waveform, and the full parameter space considered in our analysis is: \(\vec{\theta} = \{M, \eta, D_L, t_c, \phi_c, \chi_1, \chi_2, \iota, \bar{\alpha}_{\rm dCS}\}\). Here \(D_L\) is the luminosity distance and \(\iota\) is the inclination angle, which is defined as the angle between the line of sight of detector and the direction of angular momentum of the binary. 
\(\Gamma^{-1}\) is the covariance matrix, given by the inverse of the FIM \(\Gamma_{ab}\) \cite{Finn:1992, Cutler:1994}. The FIM is defined by the inner product
\begin{equation}
	\Gamma_{ab} = \left( \frac{\partial h}{\partial \theta^a} | \frac{\partial h}{\partial \theta^b} \right).
	\label{FIM}
\end{equation}
For any pair of frequency-domain signals \(p(f)\) and \(q(f)\), the inner product appearing in Eq.~\eqref{FIM} is defined as
\begin{equation}
	(p|q) \equiv 2 \int_{f_{\rm low}}^{f_{\rm high}} \frac{p^(f)\,q(f) + p(f)\,q^(f)}{S_n(f)} \, df \,,
	\label{inner product}
\end{equation}
where \(S_n(f)\) denotes the one-sided power spectral density of the detector. The integration band—the low- and high-frequency cutoffs—are set by the physical characteristics of the waveform. Specifically, the upper cutoff is taken as the frequency at the innermost stable circular orbit,
\begin{equation}
f_{\rm high} = f_{\rm ISCO} = \frac{1}{6^{3/2} \pi M} \,,
	\label{eq:flow}
\end{equation}
while the lower cutoff is determined by the total observation duration \(T_{\rm ob}\):
\begin{equation}
	f_{\rm low} = \frac{1}{8\pi \eta^{3/5} M} \left( \frac{5 \eta^{3/5} M}{T_{\rm ob}} \right)^{3/8} \,.
	\label{eq:fhigh}
\end{equation}
In this work, \(T_{\rm ob}\) is defined as the time elapsed from the start of observation until the binary reaches the innermost stable circular orbit.

When a GW signal is observed simultaneously by multiple detectors, the total Fisher matrix is the sum of the individual contributions:
\begin{equation}
	\Gamma_{ab}^{\rm total} = \Gamma_{ab}^{(1)} + \Gamma_{ab}^{(2)} + \cdots \,,
\end{equation}
where the superscripts \((1), (2), \dots\) label the different detectors in the network.

\begin{figure}
	\centering
	\includegraphics[width=0.48\textwidth]{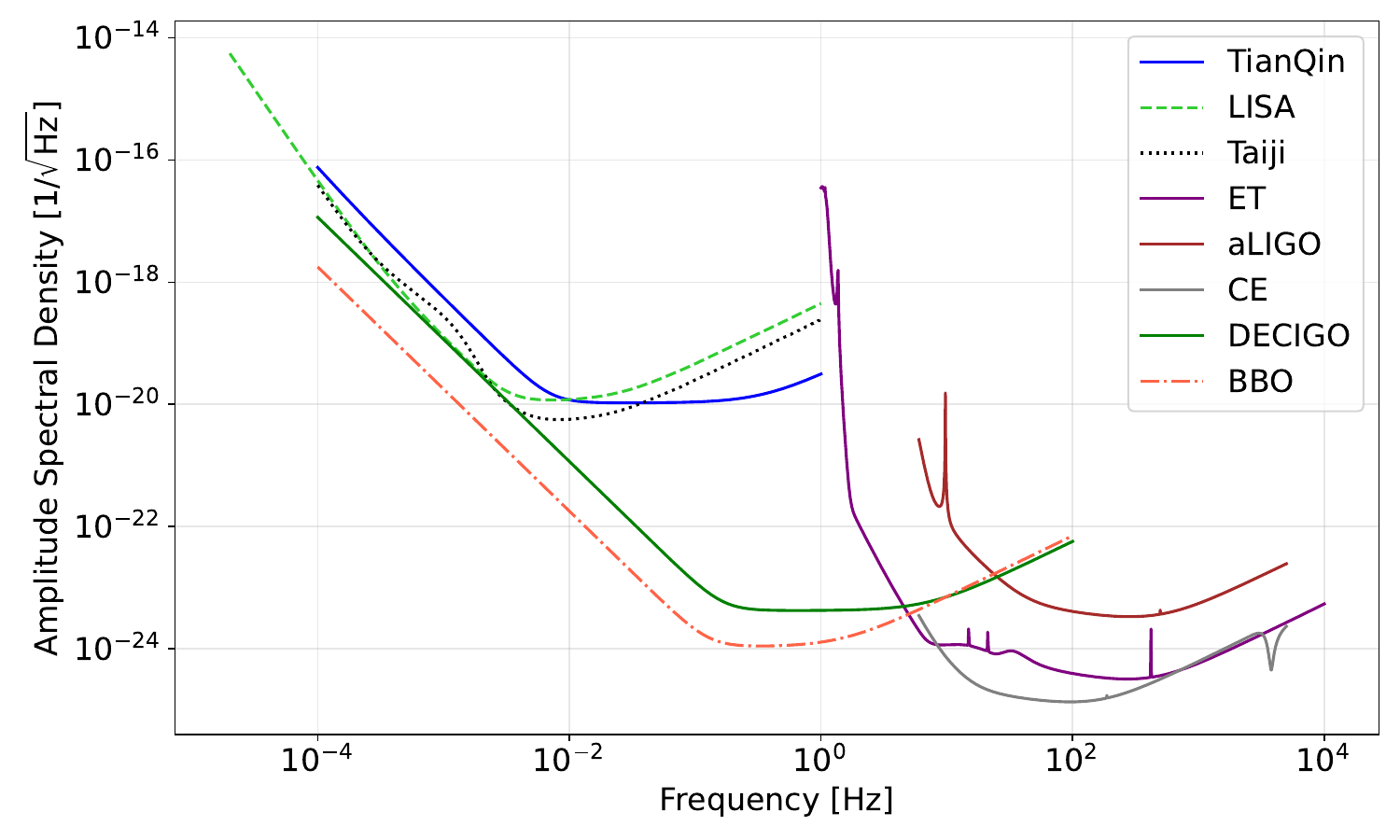}
	\caption{Sensitivities of various detectors.}
	\label{fig:sensitivity}
\end{figure}

In this paper, we consider a suite of current and future GW detectors, including third-generation ground-based observatories—such as \ac{ET} \cite{Punturo:2010zz,ET:2025xjr} and \ac{CE} \cite{Reitze:2019iox,Evans:2021gyd}—space-based missions like TianQin \cite{Luo:2015,Luo:2025ewp,Li:2024rnk}, LISA \cite{Audley:2017drz,Robson:2018ifk}, and Taiji \cite{Hu:2017mde}, as well as sub-Hertz detectors such as DECIGO \cite{Kawamura:2006up} and \ac{BBO} \cite{Harry:2006fi}. These instruments span distinct sensitivity bands and achieve varying levels of strain sensitivity. Fig. \ref{fig:sensitivity} provides a schematic overview of their respective sensitivity curves.

\section{Results}
\label{sec:result}
We now present projected constraints on dCS gravity from future gravitational-wave detectors. Our analysis first quantifies how measurement precision depends on binary source parameters, then highlights the gains from multi-band observations of stellar-mass black hole binaries—combining third-generation ground-based and space-based detectors. Finally, using realistic astrophysical population models, we estimate the expected upper limits on the dCS coupling scale achievable with upcoming observatories.

\begin{figure*}
	\centering
	\includegraphics[width=0.95\textwidth]{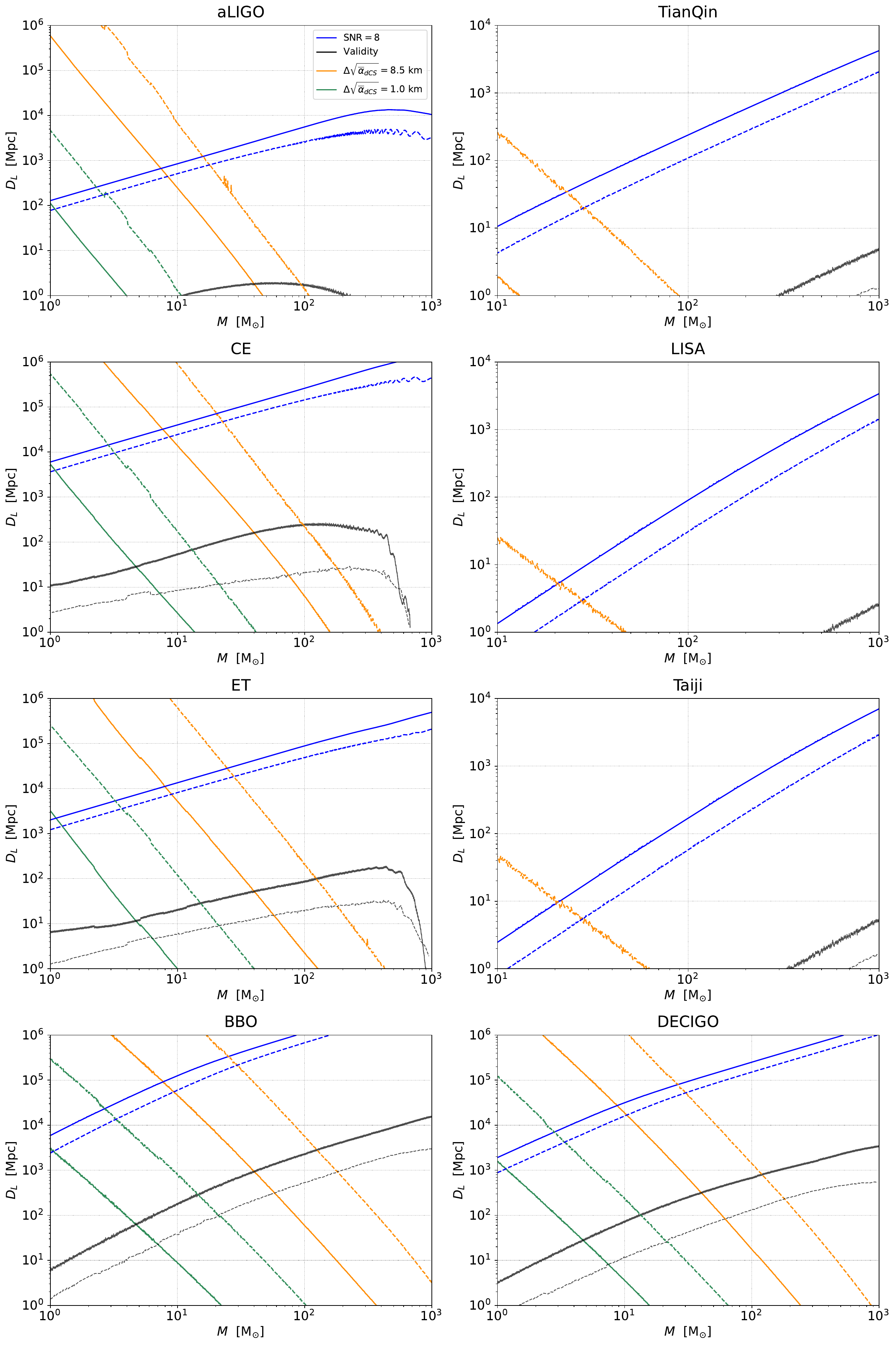}
	\caption{Dependence of  constraint on $\sqrt{\bar{\alpha}_{\rm dCS}}$ on $M$ for various GW detectors, with stellar mass black holes.}
	\label{fig:massall}
\end{figure*}

The sensitivity of \ac{GW} detectors to \ac{dCS} gravity strongly depends on the total mass of the binary system. On one hand, the total mass significantly influences the phase deviation in the gravitational waveform induced by dCS corrections (see Eq. \eqref{eq:ppedcs}); on the other hand, it directly determines the frequency band of signal, while detector sensitivity varies dramatically across frequencies.
Figure \ref{fig:massall} illustrates how the constraints on the dCS coupling parameter from various detectors evolve with the total mass of binary. Results are shown for two representative mass ratios: \( q = 1.2 \) (solid line), corresponding to GW150914—the first detected GW event—and \( q = 8.9 \) (dashed line), matching GW190814, which has the highest mass ratio observed to date.
The blue curves indicate the detection horizon distances at a \ac{SNR} of 8. The black curves mark the horizon distances limited by the validity of the waveform model. The green and red curves show the distances at which the measurement precision reaches \(\sqrt{\bar{\alpha}_{\rm dCS}} < 8.5~\text{km}\) and \(1~\text{km}\), respectively.

The results show that ground-based GW detectors, i.e. aLIGO, CE, and ET, exhibit a decline in their ability to constrain the dCS coupling parameter for binary systems with total masses exceeding several hundred solar masses. This degradation comes from two factors: limited low-frequency sensitivity and the high-frequency cutoff imposed by the innermost stable circular orbit in the integration range used for Fisher matrix analysis. The precise turnover mass varies among detectors due to differences in their low-frequency performance.
In contrast, the trend in SNR behaves differently. SNR calculations for ground-based detectors integrate over the full inspiral–merger–ringdown waveform and typically employ a high-frequency cutoff far beyond the ISCO frequency used in the Fisher analysis, thereby mitigating the impact of the ISCO truncation on detectability.

Furthermore, systems with lower total masses and higher mass ratios generally yield tighter constraints on the dCS coupling parameter. Although larger mass ratios tend to improve estimation precision compared to more symmetric binaries, the region of source parameter space where the perturbative waveform model remains valid is significantly narrower for high mass ratio systems. This narrowing arises because, at fixed total mass, a larger mass ratio corresponds to a smaller secondary black hole mass; according to the small coupling validity condition in Eq.~\eqref{eq:validity}, this reduces the allowable range of the dCS coupling parameter for which the waveform model is physically meaningful. By comparison, the direct effect of the mass ratio itself on parameter estimation accuracy is relatively modest.

Furthermore, due to waveform validity constraints, space-based detectors such as TianQin, LISA, and Taiji are unlikely to place meaningful constraints on dCS gravity from individual events. In contrast, while aLIGO is limited by its sensitivity, third-generation ground-based detectors, such as CE and ET, are expected to directly constrain dCS gravity through precise measurements of the GW phase. Even more promising are sub-Hertz detectors like DECIGO and \ac{BBO}, whose enhanced sensitivity could push constraints on the dCS coupling scale below the kilometer level.

\begin{figure*}
	\centering
	\includegraphics[width=0.95\textwidth]{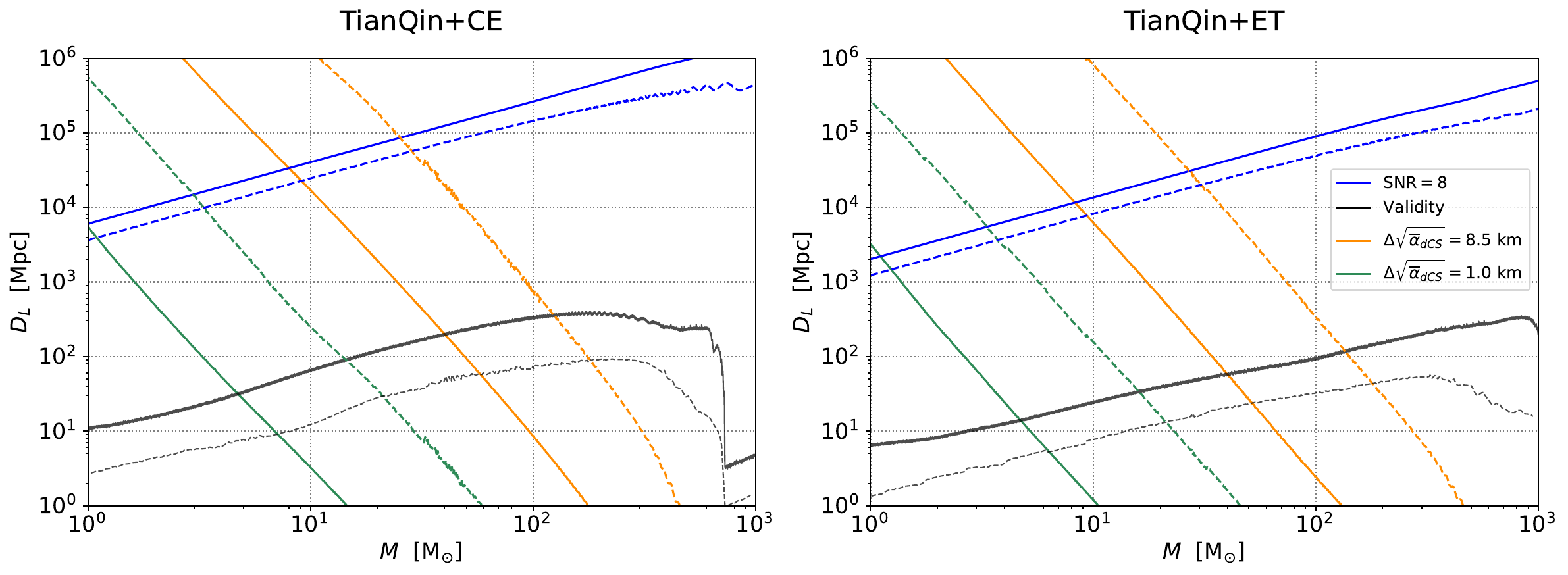}
	\caption{Dependence of constraint on $\sqrt{\bar{\alpha}_{\rm dCS}}$ on $M$ for multiband observation, with stellar mass black holes.}
	\label{fig:multiband}
\end{figure*}

For the source mass range considered in this work, joint observations by space- and ground-based detectors will enable multi-band GW astronomy. We evaluate the capacities of such joint detections—specifically, TianQin combined with either CE or ET—as shown in Fig.~\ref{fig:multiband}. The results indicate that, although space-based detectors alone struggle to impose stringent limits on dCS gravity, multi-band observations can moderately enhance overall constraining capability (a conclusion also noted in \cite{Shi:2022qno}). More significantly, multi-band observations extend the horizon distance over which the waveform model remains valid by a factor of approximately 2–3, substantially increasing the accessible population of viable sources.

Thanks to advances in both observational data and theoretical modeling, our understanding of the population properties of stellar mass black holes has significantly improved. In this work, we adopt a widely used population model for stellar mass binary black holes—the so-called BROKEN POWER LAW + 2 PEAKS model—to generate 500 simulated catalogs of binary black hole mergers.
Assuming that all detectors operate for one year, we estimate:
(i) the expected number of events that satisfy the weak-coupling condition and are thus suitable for testing dCS gravity;
(ii) the number of events per year that both satisfy the weak-coupling condition and yield constraints on the dCS coupling parameter tighter than the current best bound; and
(iii) the most stringent limits on the dCS coupling parameter achievable using stellar-mass binary black hole systems.
The results are summarized in Table~\ref{tab:catalog}.

Our analysis shows that third-generation ground-based detectors such as CE and ET are expected to yield only a modest number of dCS-testable events—on the order of \(\mathcal{O}(1)\). In stark contrast, sub-Hertz detectors like DECIGO and BBO are expected to detect many more such events—approximately \(\mathcal{O}(10)\sim \mathcal{O}(10^2)\)—and could constrain the dCS coupling parameter with a precision approaching the kilometer scale.

\begin{table*}
	\centering
	\caption{The expected detection number of events that suitable for testing dCS gravity, capable of providing constraints tighter than 8.5 km, and the most stringent constraints on the dCS theory achievable with stellar-mass binaries, of various GW detector configurations.}
	\renewcommand{\arraystretch}{1.5}
	\begin{tabular}{|c|c|c|c|c|}
		\hline
		Detector configurations& $N_{\rm \rho>8}$ &$N_{\rm \sqrt{\bar{\alpha}_{\rm dCS}}\lesssim \eqref{eq:validity}}$ & $N_{\rm \sqrt{\bar{\alpha}_{\rm dCS}}\lesssim 8.5 km}$ & $\sqrt{\bar{\alpha}_{\rm dCS}}$ (km) \\
		\hline
		CE & 3411 & 1.6 & 0.8 & 3.2\\ \hline
		CE+TianQin             & 3412 & 2.8 &1.2& 3.2\\ \hline
		ET             & 2974 & 0.4 & 0.2 & 3.2\\ \hline
		ET+TianQin            & 2975 & 0.4 & 0.2 & 3.2 \\ \hline
		DECIGO             & 3391  & 20.8 & 15.6 & 3.9 \\ \hline
		BBO        & 3688 & 507 & 319 & 2.7\\ \hline
	\end{tabular}
	\label{tab:catalog}
\end{table*}

\section{Conclusion}
\label{sec:conclusion}
In this paper, we have analyzed the capabilities of a broad suite of current and future GW detectors—including the operational ground-based detector aLIGO, third-generation observatories CE and ET, space-based missions TianQin, LISA, and Taiji, as well as sub-Hertz concepts such as DECIGO and BBO, to constrain the coupling parameter of dCS gravity using phase measurements from the inspiral signals of binary black hole systems.

Our results show that, due to limitations imposed by waveform validity, space-based detectors are unlikely to place meaningful constraints on dCS gravity. However, in the context of multi-band observations of stellar mass binary black holes by combining space- and ground-based detectors, although such joint observations offer only modest improvement in constraint precision compared to CE or ET alone, they significantly extend the horizon distance for sources suitable for dCS tests, thereby increasing the expected number of detectable events usable for testing the theory. Furthermore, by adopting the widely used Power-peak population model for stellar mass binary black holes, we performed a more detailed assessment of the capacities. The results indicate that sub-Hertz detectors like DECIGO and BBO could potentially constrain the dCS coupling parameter to a precision approaching the kilometer scale.

Nevertheless, this work has several limitations. First, the waveform corrections employed here are restricted to the inspiral phase; information from the merger and ringdown phases is not incorporated. Recent progress in computing quasi-normal modes in modified gravity theories offers a promising path toward addressing this gap in future studies. Second, our waveform corrections are limited to the leading-order contribution from quasi-circular orbits; more complete waveform models that incorporate higher-order post-Newtonian effects, spin-induced corrections, or eccentricity could reveal additional signatures and tighten constraints on the theory. Finally, all forecasts in this paper rely solely on Fisher information matrix estimates, which provide only approximate, order-of-magnitude guidance; robust and precise inference will ultimately require full Bayesian parameter estimation.

\begin{acknowledgments}
The authors thank Jianwei Mei and Zhao Li for useful discussions. This work has been supported  by the Natural Science Foundation of China (Grants  No.12405080) and
the National Key Research and Development Program of
China (Grant No. 2023YFC2206702).
\end{acknowledgments}

\bibliography{reference}

\end{document}